\documentclass[aps, prb, twocolumn, reprint, superscriptaddress, floatfix, citeautoscript,longbibliography]{revtex4-2}

\pdfoutput=1
\usepackage{graphicx}
\usepackage{siunitx}
\usepackage{dcolumn} 
\usepackage{bm} 
\usepackage{chemformula} 
\usepackage{comment}
\usepackage{braket}
\usepackage{amssymb}
\usepackage{amsmath}
\usepackage[normalem]{ulem}

\usepackage{color}

\usepackage[
   unicode,
   colorlinks = true,
   allcolors = blue,
]{hyperref}

\usepackage{cleveref}

\usepackage{natbib}

\begin{document} 
\newcommand{\etal}{{\it et al.}}
\newcommand{\pw}[1]{\SI[per-mode = symbol]{#1}{\mJ\per\square\cm}}
\newcommand{\cd}[1]{\SI[]{#1}{\per\cubic\cm}}
\newcommand{\dC}{$^{\circ}$C}
\newcommand{\Fig}[1]{{\mbox{Fig.~#1}}}
\newcommand{\Table}[1]{{\mbox{Table~#1}}}
\newcommand{\rom}[1]{\uppercase\expandafter{\romannumeral #1\relax}}
\newcommand{\Ag}{A$_{\rm{1}}$}
\newcommand{\Eg}{E$^1$$_{\rm{TO}}$}
\newcommand{\TeN}{Te N$_{4,5}$}
\title{Coherently Coupled Carrier and Phonon Dynamics in Elemental Tellurium Probed by XUV Transient Absorption}

\author{Jonah R. Adelman}
\email[Email: ]{jradelman@berkeley.edu}
\affiliation{Department of Chemistry, University of California, Berkeley, CA 94720, USA}

\author{Hugo Laurell}
\affiliation{Department of Chemistry, University of California, Berkeley, CA 94720, USA}
\affiliation{Material Sciences Division, Lawrence Berkeley National Laboratory, Berkeley, CA 94720, USA}
\affiliation{Department of Physics, Lund University, Box 118, 22100 Lund, Sweden}

\author{Lauren B. Drescher}
\altaffiliation{Current address: Max-Born-Institut, Max-Born-Straße 2A, 12489 Berlin, Germany}
\affiliation{Department of Chemistry, University of California, Berkeley, CA 94720, USA}
\affiliation{Chemical Sciences Division, Lawrence Berkeley National Laboratory, Berkeley, CA 94720, USA}
\affiliation{Department of Physics, University of California, Berkeley, CA 94720, USA}

\author{Han K. D. Le}
\affiliation{Department of Chemistry, University of California, Berkeley, CA 94720, USA}
\affiliation{Material Sciences Division, Lawrence Berkeley National Laboratory, Berkeley, CA 94720, USA}

\author{Peidong Yang}
\affiliation{Department of Chemistry, University of California, Berkeley, CA 94720, USA}
\affiliation{Material Sciences Division, Lawrence Berkeley National Laboratory, Berkeley, CA 94720, USA}
\affiliation{Department of Materials Science and Engineering, University of California, Berkeley, CA 94720, USA}
\affiliation{Kavli Energy NanoScience Institute, Berkeley, California 94720, United States}

\author{Stephen R. Leone} 
\email[Email: ]{srl@berkeley.edu}
\affiliation{Department of Chemistry, University of California, Berkeley, CA 94720, USA}
\affiliation{Chemical Sciences Division, Lawrence Berkeley National Laboratory, Berkeley, CA 94720, USA}
\affiliation{Department of Physics, University of California, Berkeley, CA 94720, USA}

\date{\today} 

\begin{abstract}

The narrow bandgap semiconductor elemental tellurium (Te) has a unique electronic structure due to strong spin-orbit splitting and a lack of inversion symmetry of it's helical lattice.  Using broadband extreme ultraviolet core-level transient absorption, we measure simultaneously the coherently coupled photo-induced carrier and lattice dynamics at the \TeN{} edge initiated by a few-cycle NIR pulse. Ultrafast excitation of carriers leads to a coherently excited \Ag{} phonon oscillation and the generation of a hot carrier population distribution that oscillates in temperature, and the phonon excursion and hot carrier temperature are $\pi$ out of phase with respect to each other. The depths of modulation suggest a significant coupling between the electronic and lattice degrees of freedom in Te. A long-lived shift of the absorption edge suggests an excited state of Te in a new equilibrium potential energy surface that lives on the order of the carrier recombination timescale. The observed phonon-induced oscillations of the hot carriers are supportive of a change in the metallicity, whereby Te becomes more metallic with increasing phonon-induced displacement. Additionally, near the Fermi level we observe an energy-dependent phase of the displacive excitation of the \Ag{} phonon mode. The discovery of coherent coupling between the lattice and hot carriers in Te provides the basis to investigate coherent interactions between spin and orbital degrees of freedom. The results spectrally and temporally resolve the correlation between photo-excited hot carriers and coherent lattice excitations, providing insight on the optical manipulation of the Te electronic structure at high carrier densities exceeding \cd{d21}. 

\end{abstract}

\maketitle 

\section{Introduction} 
\label{intro}
Tellurium (Te) is an elemental semiconductor that has attracted extensive interest due to its chiral structure of connected Van der Waals helical chains \cite{Bradley1924} (\Fig{\ref{figure:TA_map}}(a)). The strong spin-orbit interaction and chiral structure lead to many exotic electronic properties that include radial spin textures \cite{Sakano2020}, topological phase transitions \cite{Ideue2019}, and a quantum Hall effect \cite{Qiu2020}. An exceptionally high field effect mobility \cite{Kim2022}, resistance to oxidation \cite{Kim2022}, and thickness-tunable band gap of Te nanowires enables potential application in p-type transistors at the quantum confinement scale \cite{Kramer2020}. The high level of nested electronic bands also inherently leads to favorable thermoelectric properties in p-type Te where knowledge of carrier-phonon scattering is of vital importance \cite{Lin2016}. Underlying these favorable properties are two anisotropic lone electron pairs at the valence band maximum (VBM) and isotropic anti-bonding $5p$-orbitals at the conduction band minimum (CBM). Important to realizing applications with Te is understanding the non-equilibrium carrier dynamics in both its valence and conduction bands. Manipulation of the chiral structure is possible by excitation of the electrons, which thus enables coupling of the electronic, lattice, and spin degrees of freedom. This is achievable via intense ultrafast laser pulses that may induce dynamic coupling between the electrons and lattice \cite{Ning2022}. While optical-pump probe \cite{KamarajuPRB2010,Ning2022,Roeser2002} and photoelectron spectroscopy \cite{Lyu2023} have made progress in the measurement of lattice and carrier dynamics in Te, simultaneous measurement remains difficult. In these regards, a time-resolved spectroscopic method that can resolve both the hot carriers and excited lattice configuration will be helpful in understanding the coupling of the lone pair and anti-bonding orbitals to obtain useful properties.

In previous ultrafast studies of elemental Te, laser excitation was shown to induce a transient semiconductor-to-metal transition by distortion along the \Ag{} phonon coordinate \cite{Ning2022,Roeser2002,Kim2003}. The distortion introduced by the \Ag{} phonon motion is depicted \Fig{\ref{figure:TA_map}}(b,c). It was also demonstrated, using polarized optical transient reflectivity spectroscopy, that there is a band-splitting in the conduction band minimum that emerges at the H-valley and decays over 30\,ps by relaxation of the carriers to the ground state \cite{Jnawali2020}. More recently, it was determined with transient reflectivity in the mid-IR spectral region that the Auger recombination at the H-valley occurs over the same timescale \cite{ZhuoPRB2024}. Coherent dynamics of the \Ag{} phonon have been studied extensively \cite{ZeigerPRB1992}, investigating the impacts of carrier density, diffusion \cite{KamarajuPRB2010}, and electronic softening \cite{Hunsche1995} of the mode. Further investigation is needed as to how the \Ag{} coherent phonon motion couples to the hot carriers that influence these effects, and what role they may have in the light-induced renormalization of the electronic structure. Additionally, discerning the energetic coupling between the hot carriers and coherent phonon is essential to clarifying the origin of the prominent coherent phonon motion in Te, where it is unclear as to whether the electronic temperature or carrier population is the driving force \cite{ZeigerPRB1992}.

In this work, we provide experimental characterization of the hot carrier dynamics and their coupling to the lattice via broadband extreme ultraviolet (XUV) transient absorption measurements at the \TeN{} edge of Te thin films. The core-level transitions from the Te $4d$ shell to the Fermi level probe the carriers in the valence band structure, changes to the lattice, and transient modification of the electronic structure with few-femtosecond resolution. An example of possible pump and probe transitions in our experiment are shown in \Fig{\ref{figure:TA_map}}(d) over a portion of the band structure. We: (1) track the carrier thermalization and cooling processes, (2) characterize coherent coupling of the lattice and electronic degrees of freedom, and (3) measure the photo-induced lattice distortion coupled to the electronic structure. By exciting Te with a few-femtosecond laser pulse that spans visible to NIR wavelengths (500-1000\,nm), at carrier densities around \cd{d21}, we observe the generation of hot holes and electrons in the electronic band structure and coherent excitation of the \Ag{} phonon mode. The high energy carriers rapidly thermalize through electron-electron and electron-phonon (196(8)\,fs) scattering to a hot Fermi-Dirac distribution and subsequently cool over the course of 1.59(3)\,ps. Due to coupling between electronic and lattice degrees of freedom, excitation of the \Ag{} mode coherently modulates the temperature of the hot carriers with an observed nearly $\pi$ phase difference between the hot carriers the phonon-driven displacement. We attribute this to a variation of the electronic heat capacity as Te becomes more metallic when the structure is symmetrized by the \Ag{} mode. This coherent anti-correlation between phonon excursion and carrier temperature, predicted by Giret \etal{} \cite{Giret2011} and previously observed in the A7 semimetals \cite{Geneaux2021,Drescher2023} and CDW materials \cite{ZhangPNAS2020}, is also now observed in a degenerate semiconductor. Further, we measure a long-lived distortion of the lattice along the \Ag{} phonon mode coordinate that decays over 33(5)\,ps. This is consistent with the hypothesis that carrier recombination times in Te \cite{ZhuoPRB2024} determine the lifetime of the excited lattice configuration. Lastly, we resolve the spectral phase of the coherent phonon, which provides insight into the effects of hot carrier relaxation in the generation of the \Ag{} phonon.

\section{Experimental Setup}

In the experiments we used polycrystalline Te films that are electron beam evaporated (Lebow) onto 30\,nm thick silicon nitride windows (Norcada). The Te thickness was confirmed by atomic force scanning tip microscopy to be approximately 100\,nm (see \Fig{\ref{figure:AFM}} in Appendix B). The trigonal form of Te is expected to form following crystallization of initially amorphous films over the course of a few minutes at room temperature \cite{Zhao2021}. This was confirmed by x-ray diffraction of the Te samples on silicon nitride (see \Fig{\ref{figure:XRD}} in Appendix B). To avoid sample evaporation, ablation, or laser pulse induced amorphization, careful consideration was taken to reduce laser induced heating during the experiment by chopping both the pump and probe to a repetition rate of 100\,Hz. The XUV pump-probe experiments were performed at room temperature with a broadband attosecond XUV probe generated by focusing a near single cycle visible to near infrared laser pulse into krypton gas for High Harmonic Generation (HHG) \cite{Ferray1988}. The same optical laser pulse is split before the HHG generation region in a Mach-Zender interferometer where it is delayed in time relative to the XUV probe from -100 to 3000\,fs in steps of 6.6\,fs. A measurement at shorter timescales used 0.66\,fs steps from -30 to 90\,fs along with parallel XUV transient absorption measurements in neon to obtain the zero-delay and the instrument response of 6.4(5)\,fs (see Appendix \ref{IRF}). The carrier density induced by the pump is calculated by using the linear absorption coefficient and reflectance at 800\,nm \cite{Tutihasi1969}. In the sample, the pump has an estimated peak power density of {\SI[]{1.2d12}{\watt\per\square\cm} and corresponding field strength of {\SI[]{0.21}{\volt\per\angstrom}. Further details on the instrument and XUV source are in Appendix \ref{instrument}.

\label{measurements}

\section{Results} 

In this section we first present the XUV transient absorption spectra at an estimated 1.5 $\cdot$ \cd{d21} carrier density. Subsequently, the spectral contributions are decomposed to disentangle the electronic and lattice degrees of freedom. We use an analytical model and a bi-linear decomposition for analysis. Finally, the temporal response of the spectral contributions are characterized and fitted with a multi-temperature model. 

\subsection{ XUV Transient Absorption}

\begin{figure*}[!htb]
\begin{center}
\includegraphics[width=\textwidth]{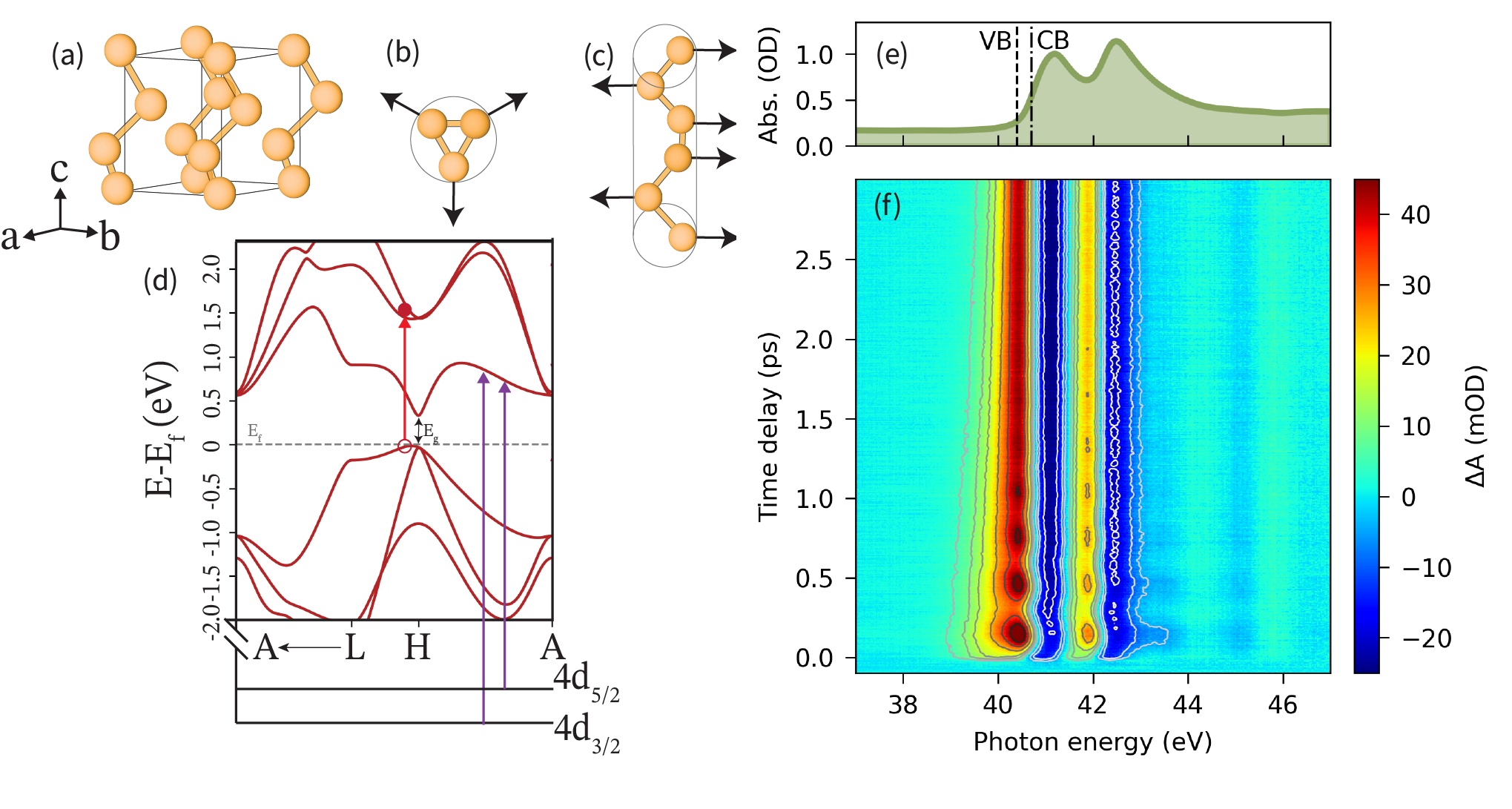}
\caption{(a) The crystal structure of elemental Te showing one of the possible chiral space groups ($P3_121$). The \Ag{} phonon motion projected into two dimensions as black arrows for a single chain viewing top down along the $c$-axis (b) and from a sideview (c). The radius dimension is used to quantify the motion. (d) A non-spin-polarized band structure of Te in vicinity of the band gap along the A-L-H-A path of the Brillouin zone, where the bandgap ($\rm{E_g}$) is located at the H-valley and nominal Fermi level ($\rm{E_f}$) is the zero reference \cite{Jain2013}. One possible valence to conduction band transition among many accessible by the few-cycle NIR pump is denoted as a red arrow. The purple arrows from the Te $4\rm{d}_{5/2}$ and $4\rm{d}_{3/2}$ semi-core to conduction band are representative transitions that compose the ground state absorption spectrum in the absence of pump-induced modifications. (e) Ground state XUV absorption spectrum. The VBM and CBM from the $4\rm{d}_{5/2}$ core-level are denoted as dashed lines and spaced by the 330\,meV bandgap. (f) Measured XUV transient absorption spectra at the \TeN{} edge between -100 and 3000\,fs at a carrier density of 1.5 $\cdot$ \cd{d21}. The time axis corresponds to the arrival of the XUV probe relative to the optical pump arriving at time zero. Photoexcitation of Te with the pump launches a coherent phonon motion that appears as the observed oscillation pattern in the differential absorption.}
\label{figure:TA_map} 
\end{center}
\end{figure*}

In \Fig{\ref{figure:TA_map}}(e), the ground state core-absorption spectrum at the \TeN{} edge in the range of 37 to 47\,eV is shown. The increase in absorption, corresponds to the $4\rm{d}_{3/2}$ (42.3\,eV) and $4\rm{d}_{5/2}$ (40.7\,eV) semi-core electrons excited to the conduction band (CB) edge. The well resolved spin-orbit doublet with a 1.6\,eV splitting and smooth absorption at higher energies is in agreement with synchrotron measurements of Te \cite{Givens1955}. The absorption onset ($<$ 41\,eV) would correspond to transitions from $4\rm{d}_{5/2}$ to the unoccupied $5p$-orbitals at the H-point followed by higher lying transitions across the entire Brillouin zone (BZ). At energies above the $4\rm{d}_{3/2}$ peak there are small features that correspond to forbidden transitions to $5d$-bands well above the conduction band minimum (CBM) \cite{Sonntag1973}. Due to the strong core-hole screening at the \TeN{} edge, as previously reported \cite{Attar2020}, the absorption structure should well reflect a core-hole modified density of states (DOS) in the conduction band. The lifetime broadening from the core-hole (on the order of 0.5\,eV \cite{Sonntag1973}), however, precludes our ability to resolve specific valleys in the absorption. The position of the valence band maximum in Te from the $4\rm{d}_{5/2}$ is previously reported to be 40.4\,eV \cite{Fuggle1980} and in p-type Te, the Fermi level is pinned near the valence band maximum (VBM) \cite{Dasika2021}. We expect our films to be naturally p-type, typical of evaporated Te \cite{Zhao2021}. 

To track the dynamics, the change in the core-level absorption from the ground state at the \TeN{} edge is detected by measuring the transmitted XUV with and without a time-delayed optical pump pulse. The XUV transient absorption spectra are shown in \Fig{\ref{figure:TA_map}}(f) at a carrier density of 1.5 $\cdot$ \cd{d21}. Typical of XUV transient absorption of thin films, a delay independent background before time zero that is due to static heating is subtracted (see details in Appendix \ref{instrument}). Following the arrival of the pump pulse at time zero, there is an increase in absorption starting at 39\,eV, which then decreases around 41\,eV. This pattern is replicated by the spin-orbit doublet and persists beyond our measured time delay. Within the first picosecond, the oscillation pattern centered around 40.5\,eV and 42\,eV resembles a shifting back and forth of the absorption edge at a frequency of 3.15\,THz. The oscillation pattern is also visible above 43\,eV in the $5d$ bands far from the Fermi level. The oscillation frequency of the absorption edge is consistent with the coherent excitation of the \Ag{} phonon, but shifted from the equilibrium frequency of 3.6\,THz \cite{Jnawali2020}. Furthermore, at 40.5\,eV, the \Ag{} phonon oscillation exhibits a tilt in time with respect to energy, indicative of a non-flat phase distribution of the coherent motion versus photon energy.

In addition to the lattice dynamics shifting the absorption edge, an increase in absorption is present below the VBM at 40.4\,eV. This may be attributed to effects of state blocking/filling, resulting from the transitions of the $4\rm{d}_{5/2}$ core level to newly unoccupied states in the VB \cite{Attar2020}. We therefore attribute this transient signal to the photo-generated hot holes in the VB. At the earliest timescale of the measurement, the increase in absorption would at most extend down to 38.2\,eV, but will narrow in time as high energy (hot) holes relax (See Appendix \ref{Appendix:Residual}). Compared to the holes, which are spectrally isolated below the VBM of the $4\rm{d}_{5/2}$, a decrease in absorption above the Fermi level from hot electrons is not as readily discerned due to spectral congestion between the lattice and carrier features. In the next section we focus on isolating the spectral contributions of the \Ag{} phonon mode and the hot carriers.

\subsection{Decomposing the Spectral Contributions of the \Ag{} Phonon and Carriers}
\label{section:spectral}

To evaluate our observations in the TA spectra, a singular value decomposition (SVD) is utilized to decompose the distinct transient contributions with unique spectral and kinetic profiles. To validate this approach, an analytical model is presented in the Appendix \ref{Appendix:decomp} where the TA can be decomposed into an energy shift of the absorption edge, inhomogeneous broadening, and state filling/blocking \cite{Zurch2017}. The SVD approach enables isolation of the basic contributions to the TA without presumptive modeling or restriction in the spectral and temporal domain. Further, it provides noise reduction and can isolate spectral contributions arising from strongly overlapping phonon and carrier signals of the spin-orbit doublet splitting. The SVD is implemented by adding the static absorption from \Fig{\ref{figure:TA_map}}(e) to the $\Delta$A at every time point, forcing the zero order basis vector to be the non-time-dependent ground state absorption.

We find the entire transient signal is largely reproduced by three singular value components consistent with the analytical model. The spectra of the SVD are shown in \Fig{\ref{figure:spectrum1}}(b,c,d) where the spectral SVD vectors are added back to the XUV ground state absorption for comparison.  The largest SVD component (\Fig{\ref{figure:spectrum1}}(b)) comprises 56\% of the SVD decomposition and resembles a shift of the ground state absorption (\Fig{\ref{figure:spectrum1}}(a)) to lower energies. There is an additional appearance of spectral features above the $4\rm{d}_{3/2}$ peak that cannot be reproduced by shifting the ground state XUV absorption. This SVD component we assign to the absorption edge shift due to the coherent phonon, band-gap renormalization, changes to 4d core-hole potential, and displacement of the equilibrium position of Te along the \Ag{} phonon coordinate from photoexcitation. From previous ultrafast work in Te at similar carrier densities, we expect an exceptionally large magnitude shift of the Te equilibrium position on the order of 26\,pm \cite{Ning2022,KamarajuPRB2010}. This is about ten times larger than what is reported in the semimetal Bismuth (2.6\,pm) \cite{Geneaux2021}. Consistent with the anticipated large Te equilibrium position displacement, an absorption edge shift of 580(30) meV is estimated from the analytical model (see details in Appendix \ref{Appendix:decomp}). This value however should be interpreted cautiously due to approximations in deconvolving the spectrum. While many effects may shift the absorption, the most clear is the decreasing conduction band energy \cite{Ning2022,Tangney2002}. Through the \Ag{} phonon induced distortion, this decrease in conduction band energy will red shift the absorption by decreasing the transition energy from the 4d levels to the DOS. Further, the \Ag{} phonon mode and lattice 
heating will modulate the bond lengths. This influences the 4d level ionic repulsion, thus changing the 4d binding energy. The exact dependence on the \Ag{} mode requires detailed calculations of the core-level potential gradient, however, the lattice heating red shifts the absorption edge as indicated by the pre-time-zero delay heating (see \Fig{\ref{figure:raw_data}}(a)). Given the increase (decrease) in intra(inter)-chain bond distance upon increasing the chain radius, and $1/r^2$ distance scaling of the electrostatic repulsion, the core-level gradient is likely to counteract the decreasing conduction band energy with a blue shift. In addition to lattice dynamics, fast carrier screening of the 4d levels \cite{RossiNL2021}, changes in the valence to 4d level coulomb repulsion \cite{deGroot}, or changes in electron localization may also shift the edge in the absence of lattice motion \cite{Schumacher2023}.

The second largest singular value (28\% of the signal) shown in \Fig{\ref{figure:spectrum1}}(c) readily resembles an unshifted XUV ground state absorption with a large increase in absorption below the valence band maximum and suppression of the $4\rm{d}_{3/2}$ peak relative to the $4\rm{d}_{5/2}$. The increase in absorption below the VBM from the Te $4\rm{d}_{5/2}$ must be due to opening of states in the valence band (holes), while the suppression of the $4\rm{d}_{3/2}$ peak comes from electrons populating the conduction band following excitation. The state filling/blocking contributions retrieved from the analytical model agrees with the increase in absorption below the VBM in the second SVD vector as well as a decrease in the $4\rm{d}_{3/2}$ transition to the CB, respectively. Nevertheless, this singular value consistent with hot carriers incorporates both the holes and electrons. We find a small contribution of excited state broadening in the SVD decomposition (\Fig{\ref{figure:spectrum1}}(d)). Both phonons and electronic effects could give rise to broadening \cite{deGroot}, however, the relative contributions cannot be disentangled from the spectrum. 

\begin{figure}[!htb]
\begin{center}
\includegraphics[width=\columnwidth]{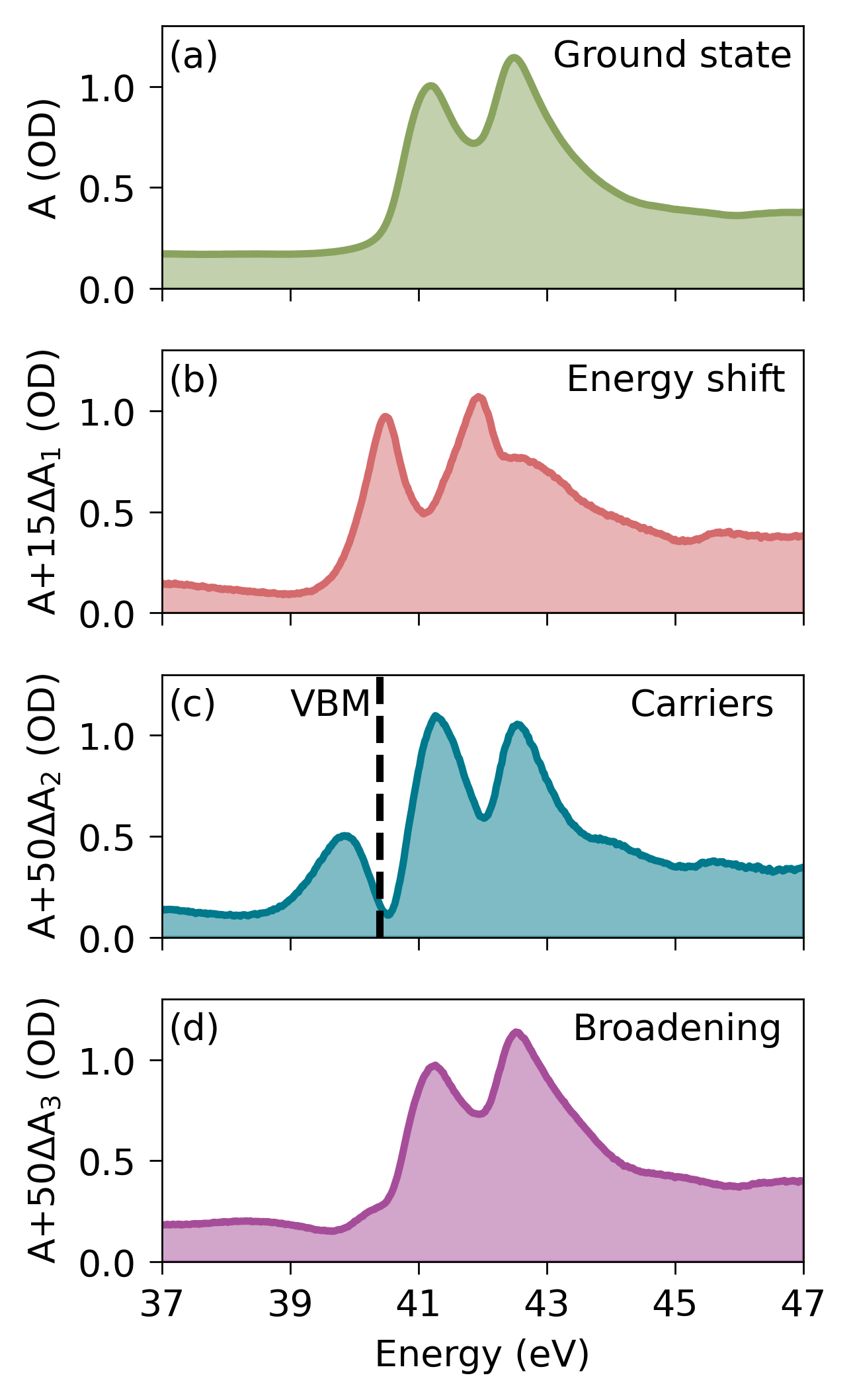}
\caption{(a) The ground state XUV absorption spectrum of elemental Te from the 4d core levels to the conduction band. Spectral contributions are retrieved from the SVD. Note the differences in the vertical scales. The shown spectra are the sum of the SVD retrieved differential absorption and ground state absorption from (a). The rescaling of the $\Delta A$ added back to the ground state XUV absorption is for visibility purposes. (b) The absorption of the excited phonon induced edge shift. (c) The absorption spectra of the hot carriers. The VBM from the  $4\rm{d}_{5/2}$ is shown as a dashed line. (d) The excited state broadening of the absorption edge.}
\label{figure:spectrum1} 
\end{center}
\end{figure}

\subsection{Temporal Dynamics of the \Ag{} Phonon and Carriers} 
\label{results_temporal_dynamics}
We now turn to the temporal dynamics of the three SVD components that track the kinetic processes following the excitation. In \Fig{\ref{figure:temporal}} the temporal vectors exhibit distinct dynamics that encode the edge shift, broadening, and carriers in detail. Focusing on the exponential rises and decays, the temporal dependence of the edge shift contribution (labeled energy shift) has a sharp initial rise within the first few femtoseconds of the optical excitation followed by a slower onset over a picosecond. The fast rise can be ascribed to a renormalization of the transitions to the conduction band typically driven by changes in the electronic screening of both the valence \cite{TranklePRB1987} and 4d core-hole \cite{RossiNL2021,Schumacher2023}. From the kink in the energy shift trace that occurs briefly after the zero-time delay, this electronic contribution can be estimated to account for nearly one-third of the energy shift. Depopulation of lone pair states near the Fermi level to anti-bonding states in the conduction band \cite{Ning2022} also yields rapid destabilization of the ground state potential energy surface resulting in a displacive force along the \Ag{} phonon coordinate that increases the helical chain radius by shifting the equilibrium position ($\Delta Q_0$). This increases the symmetry of the Te structure, inducing a displacive excitation of a coherent phonon (DECP) \cite{ZeigerPRB1992} and causes a rise on the order of half the phonon period. The phonon-induced change in local environment is what gives rise to an absorption edge shift due to changes in the 4d binding energy and decrease in the band-gap as discussed above in Section \ref{section:spectral}. The picosecond rise is typical of an incoherent growth in the acoustic phonon population that emerges from decay of the optical phonons \cite{Klemens1966}. The decay of the acoustic phonons is heavily suppressed by the very low lattice thermal conductivity of Te \cite{Lin2016} and occurs over the course of milliseconds, as determined from the presence of a large heat background in our TA spectra. 
 
\Fig{\ref{figure:temporal}} also shows the SVD component assigned to the hot carriers (blue curve). The carriers exhibit an immediate increase following excitation, but notably there is a delayed maximum. The cause of the delayed maximum is the thermalization of carriers towards the Fermi level \cite{Lyu2023}. However, the biexponential rise of the carriers suggests two potential channels feeding the thermalized carriers. The decay we attribute to a decrease in the electronic temperature of Te ($\Delta T_e$) as the hot carriers dissipate their energy through electron-phonon coupling \cite{Drescher2023}. The temporal evolution of the broadening (purple curve in \Fig{\ref{figure:temporal}}) appears with no clear decay on the measured timescale. The slow rise and lack of decay is attributable to an increase in vibrational broadening due to populating incoherent acoustic phonons, which thus increase the lattice temperature \cite{TanPRB2023}.

Notably, we observe oscillations in both the energy shift and hot carrier signals. While observation of the \Ag{} phonon in Te is well established, oscillations in the hot carrier signal at the same frequency of the phonon is similar to the back-coupling of the electronic temperature to the lattice that has been observed in semi-metallic Bi \cite{Geneaux2021}, Sb \cite{Drescher2023}, and CDW system TaSe$_2$ \cite{ZhangPNAS2020}. The hot carrier signal follows the \Ag{} phonon motion, indicating the coupling arises from movement along the \Ag{} phonon coordinate, corresponding to changes in the Te chain radius. The $\pi$ phase shift implies that when the Te helices are symmetrized (phonon maximum excursion), the carriers have a local temperature minimum. We discuss the potential source of this phase shift and quantify the temporal dynamics in the following section \ref{sec:discussion}.

\begin{figure}[!htb]
\begin{center}
\includegraphics[width=\columnwidth]{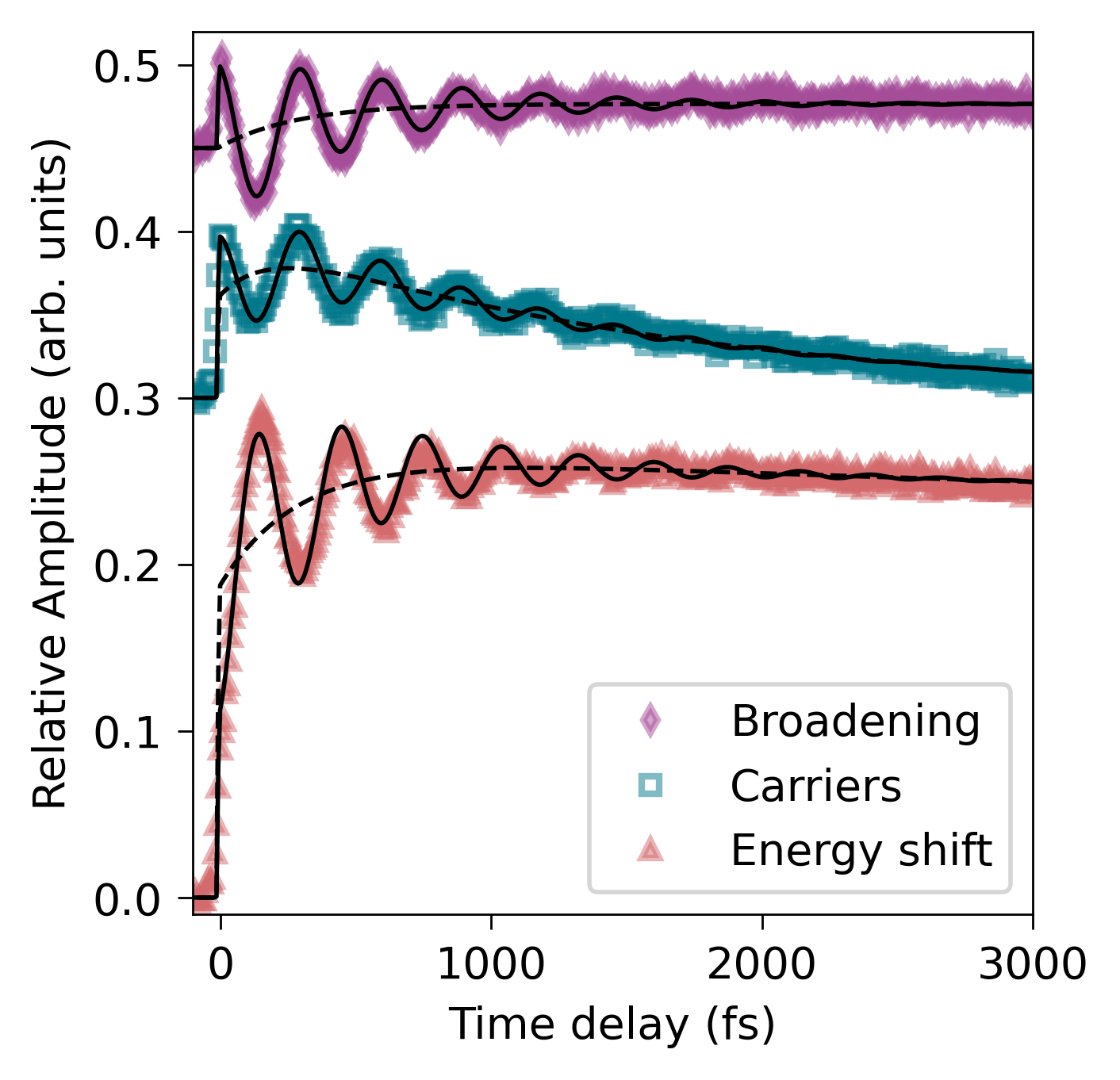}
\caption{ The temporal vectors retrieved from the SVD at a carrier density of 1.5 $\cdot$ \cd{d21}. For each component, the solid black lines are the fit to the multi-temperature model in equation \ref{eq:fit}. The dashed lines correspond to the multi-temperature model with the oscillations removed. The vectors are vertically offset for clarity.}
\label{figure:temporal} 
\end{center}
\end{figure}

\section{discussion}
\label{sec:discussion}
\subsection{Multi-Temperature Model }
\label{MTM}
To quantify the dynamics of the phonons and carriers the three SVD traces were globally fit weighted by their singular values using (Equation \ref{eq:fit}) which corresponds to a multi-temperature model. A Gaussian instrument response function, $P(t)$, is retrieved by an external measurement and fixed in the fitting routine. In this model, the hot carriers are an electronic temperature ($\Delta T_e$) bath that may transfer energy to the lattice via an electron-phonon scattering rate ($\lambda$). The electronic temperature receives energy via the initially non-thermal carriers following photo-excitation through fast $T_{ee}$ and slow $T_{ep}$ scattering channels consistent with the biexponential rise of the hot carriers (e = electronic and p = phonon). The energy shift is described by the tellurium atom displacement from the equilibrium position along the \Ag{} coordinate ($\Delta Q_0$), the acoustic phonon temperature ($\Delta T_a$), and carrier-induced changes in electronic screening ($\Delta$E). The acoustic phonon temperature is fixed to rise at twice the rate of the coherent phonon decay ($\gamma$) as the coherent optical phonon has been shown to decay via a Klemen's like mechanism \cite{Klemens1966} through an anharmonic coupling of the lattice displacement to degenerate acoustic modes \cite{Teitelbaum2018,FahyPRB2016,GarlPRB2008}. The rate for the initial rise is described by $T_r$ and relaxation back to the ground state potential energy surface ($\kappa$). The decay of $\Delta$E is shared with $\Delta Q_0$ in this model, under the approximation they both relate to the carrier population. Finally, the coherent energy exchange between the coherent optical phonon and electronic temperature is described following the energy balance equations coupled to a classical oscillator approach of Giret \etal{} \cite{Giret2011}. This includes the \Ag{} frequency ($\omega$), coherent phonon phase ($\phi$), and a frequency chirp of ($\beta$). The transient broadening $\Delta \sigma$ in Equation \ref{eq:fit} is fit and follows the acoustic phonon temperature $\Delta T_a(t)$ and phenomenologically the electronic component of the coherent energy exchange.

\begin{widetext}
\begin{equation}
\begin{aligned}
\Delta T_e(t)  &=  P(t-t_0) \circledast  [ A(1-e^{-T_{ep}t}) e^{-\lambda t} +  B(1-e^{-T_{ee}t}) e^{-\lambda t} + C\cos (\omega t + \beta t^2 +\phi +\pi )e^{- \gamma t} ]  \\
\Delta \rm{E} + \Delta Q_0(t) + \Delta T_a(t)   &=  P(t-t_0) \circledast [F(1-e^{-T_r t})e^{-\kappa t}+D(1-e^{-2\gamma t}) + E\cos (\omega t + \beta t^2 +\phi )e^{- \gamma t} ]  \\ 
\Delta \sigma(t)    &=  P(t-t_0) \circledast [G(1-e^{-2\gamma t}) + H\cos (\omega t + \beta t^2 +\phi )e^{- \gamma t} ]  \\ 
\end{aligned}
\label{eq:fit}
\end{equation} 
\end{widetext} 

At a carrier density of 1.5$\cdot$\cd{d21}, the \Ag{} phonon frequency retrieved is $\omega$ = 3.17(1)\,THz compared to the equilibrium 3.6\,THz measured by Raman spectroscopy (see black curves in \Fig{\ref{figure:temporal}}). The softening of the \Ag{} mode is not surprising as the frequency is known to be dependent upon both carrier density \cite{Hunsche1995,KamarajuPRB2010}, and temperature in the form of anharmonic phonon-phonon coupling \cite{KamarajuPRB2010}. However, when subjected to a few percent excitation of the valence electrons, an increase in temperature is expected to signifigantly blue shift the \Ag{} phonon frequency \cite{KamarajuPRB2010}. The 12 \% decrease in frequency of the \Ag{} motion measured with the XUV can then be attributed to a decrease in bond strength as the non-bonding lone pairs at the Fermi level are nearly instantaneously excited into the conduction band anti-bonding orbitals. This is in good agreement with the measured frequency determined from optical reflectivity pump-probe spectroscopy of polycrystalline Te \cite{Hunsche1995,KamarajuPRB2010}. Similarly, at a carrier density of 2 $\cdot$ \cd{d20} the \Ag{} phonon frequency is found to be 3.6(3)\,THz, closer to the equilibrium value (See Appendix \ref{appendix:low_fluence}) with no apparent decay. The coherent phonon lifetime of $1/\gamma = 0.58(1)$\,ps and chirp of $\beta =$ 0.89(7)\,ps$^{-2}$ at 1.5$\cdot$\cd{d21} is typical of high fluence \cite{KamarajuPRB2010}, while the coherent phonon phase is determined to be $\phi =$ -0.04(1)$\pi$\,rad and is discussed further below in Section \ref{sec:phonon}. 

The rise of the energy shift signal ($T_r$) and decay ($\kappa$) of the Te displacement from the equilibrium position ($\Delta Q_0$) is now considered. The lifetime of the decay of $\Delta Q_0$ from the excited state potential energy surface as determined by the multi-temperature model is $1/\kappa =  33(5)$\,ps and is much slower than the hot carrier decay. This lifetime of the Te displacement in the excited state potential energy surface is nearly identical to the carrier Auger recombination lifetime (31\,ps) at the band-edge measured via optical pump-probe spectroscopy in 100\,nm thick flakes \cite{ZhuoPRB2024} and the photo-induced degeneracy lifting of the conduction band minimum (H$_6$ band) relaxation time (30\,ps) \cite{Jnawali2020}. In the original formulation of the displacive excitation of a coherent phonon model it is unclear whether the displacement $\Delta Q_0$ is dependent upon the carrier population, $n(t)$, or if only the electronic temperature contributes to the displacement \cite{ZeigerPRB1992}. It is notable that $\Delta Q_0$ in the multi-temperature model follows a timescale comparable to Auger recombination of fully relaxed carriers, such that $\Delta Q_0(t) \propto \kappa n(t)$. This could potentially explain why the displacement of the Te equilibrium position is much larger than in A7 semimetals \cite{Drescher2023, Geneaux2021} where it is proposed to follow the electronic temperature. Additionally, when measured at a carrier density of $2\cdot$\cd{d20}, $\Delta Q_0(t)$ has no measurable decay out to the maximum time delays tested ($\sim$ 2\,ps). This is consistent with lengthening of the carrier lifetime as Auger recombination is reduced at low carrier densities, and multi-phonon mediated recombination is only expected to occur over 200\,ps \cite{ZhuoPRB2024}. Moreover, there is a fast rise ($T_r$) to the energy shift trace, which is instantaneous with the sub 5\,fs pump excitation (see Appendix \ref{section:early}). This rise, while too fast to be explained by the phonon motion, likely originates in carrier-induced electronic structure changes such as changes in electron localization \cite{Schumacher2023} or 4d core-hole screening \cite{RossiNL2021}.

The dynamics of the hot carriers are examined in \Fig{\ref{figure:temporal}}. Upon initial excitation, the non-thermal carriers far from-equilibrium will quickly redistribute their energy through various scattering channels as they thermalize. The biexponential rise of the multi-temperature model for the hot carriers includes a fast term $T_{ee}$ that closely follows the 6.4(5)\,fs instrument response, and a slower $T_{ep} = $ 196(8)\,fs rise. The latter is extracted by deconvoluting the rise dynamics from the cross-correlation, while the fast term is not readily resolved. Both timescales are much faster than previously measured inter-valley scattering that necessitates a population of high momentum acoustic phonons \cite{Jang2024,Jnawali2020}. Thus the primary source is likely intra-valley scattering towards the Fermi level in the form of energy-relaxation that is typically on the order of 100 fs \cite{Ichibayashi2009,ZLin1988}. It is however possible that both carrier-carrier and carrier-optical phonon scattering are the dominant mechanisms for thermalization \cite{Wilson2020,Attar2020}. While carrier-carrier scattering can only redistribute energy within the electronic system, the maximum energy exchange is on the order of the 1.55\,eV photon energy. This will lead to much faster thermalization than through optical phonon emission that is constrained to the tens of meV transfer per scattering event. The fast rise ($T_{ee}$) is thus attributable to carrier-carrier scattering, while the slower thermalization channel $T_{ep}$ can be attributed to electron-phonon coupling, which is similar to the 0.3 ps intra sub-band cooling time measured with photoelectron spectroscopy \cite{Lyu2023}. The discrepancy could be due to the fact that in the analysis the hot carrier signal included both holes and electrons and that there are differences in scattering rates for holes versus electrons. The hot carrier decay ($\lambda$) has a relaxation time of $1/\lambda =$1.59(3)\,ps that is similar to the timescale (1.3\,ps \cite{Jang2024}, 1.7\,ps \cite{Lyu2023}) for the transfer of high-energy carriers to the lowest energy sub-bands of the H-valley. In this process, for the hot carrier decay, the electronic temperature $\Delta T_e$ is reduced as optical-phonon emission through electron-phonon coupling that transfers energy to the lattice. The hot carriers determined from the SVD do not include the cooled lowest energy carriers that share the long lifetime of the Te equilibrium position displacement. Similarly, the non-thermal carriers are shown in the residual TA spectra alongside the cooled carriers (See \Fig{\ref{figure:resdiual}} of Appendix \ref{Appendix:Residual}), where there are residual changes in absorption near the Fermi-level that rapidly decay on a timescale of a hundred femtoseconds. This underlies a clear limitation of the analysis above. The SVD is not sensitive to early timescale temporal variations in the spectral profile (on the order of a mOD), such as the non-thermal carriers and their deviation from a Fermi-Dirac distribution in the first tens of femtoseconds. Thus, similar to other XUV TA studies, the thermalization timescales are retrieved by analyzing the appearance of a maximum in the hot carrier population, as non-thermal carriers may produce only small spectral deviations \cite{ChangPRB2021}.

\subsection{Coupled Carrier and \Ag{} Phonon Dynamics}
\label{sec:phonon}

We now consider the coherent modulations of $\Delta T_e(t)$ that are observed. Following the absorption of an ultrashort optical laser pulse by Te, there is a rapid increase in the total energy of the electronic degrees of freedom. This abrupt increase in the carrier population ($n(t)$) and electronic temperature ($\Delta T_e(t)$) contributes to excite the \Ag{} phonon mode and is well explained by the DECP model \cite{ZeigerPRB1992} and additional theory by O’Mahony \etal{} \cite{Mahony2019}. While Te is a semiconductor in the ground state, the displacement along the \Ag{} phonon coordinate leads to a symmetric Dirac metal through a light-induced phase transition \cite{Ning2022}. This phase transition can similarly be achieved by depopulating the doubly degenerate VBM lone-pair orbitals with shear strain to the CBM \cite{Ideue2019}. The increase in metallic character when Te becomes more symmetric by increasing the helical chain radius should lead to an increase in the electronic heat capacity \cite{Ashcroft1976} as the DOS is increased at the Fermi level \cite{Tangney2002}. If we consider the temporal evolution of the hot carriers in \Fig{\ref{figure:temporal}}, the phase offset is close to $\pi$ with respect to the phonon excursion. When the phonon displacement is maximum and Te most metallic, the electronic heat capacity is maximized, leading to a lower temperature of the hot carriers. This similarly follows the thermodynamic model of Giret \etal{} \cite{Giret2011} that describes a coupling between the electronic temperature and lattice motion in Bi as an isentropic conservation of entropy. The same model has been applied to interpret XUV pump-probe results in Sb \cite{Drescher2023} and Bi \cite{Geneaux2021}. As shown by Ning \etal{} \cite{Ning2022}, \Ag{} coherent phonon motion leads to vast modifications of the band structure in the vicinity of the Fermi level, and at a maximum distortion, metallization of Te. Given this, the dependence of the electronic heat capacity on $\Delta Q_0$ could exhibit large variation due to how it strongly couples to the band structure, particularly if there is a large excursion of the \Ag{} phonon mode (given as a radius). This is not surprising as there are numerous band crossings at the Fermi level that emerge upon increasingly symmetrizing Te \cite{Ning2022,Tangney2002}. While an indirect band gap is closed at the A-valley for a 14\,pm displacement \cite{Tangney2002}, complete symmetrizing of Te at 26\,pm additionally closes the gap at the H-valley \cite{Ning2022}. Although our level of analysis precludes determination of a band gap closing, or the extent of symmetrization, the oscillations of the hot carrier are indicative of a strong influence of the \Ag{} phonon on the electronic heat capacity, and thus the transiently changing number of available states in proximity to the Fermi level. Calculations of the electronic heat capacity changes that occur from the light-induced excitation of the \Ag{} phonon mode may prove useful in the future in understanding these observations, and to determine which valleys may contribute the most for the carriers excited.

Finally, we consider a Fourier analysis of the phonon-induced oscillations shown in \Fig{\ref{figure:AC_map}}. This is accomplished by subtracting the exponential contributions to the TA captured by the SVD and adding back the residual TA of \Fig{\ref{figure:resdiual}}. Further, a decomposition of the spin-orbit split 4d core-levels is done following the methodology of Zuerch \etal{} \cite{Zurch2017} to eliminate spectral congestion between the $4\rm{d}_{5/2}$ and $4\rm{d}_{3/2}$ to conduction band transitions. \Fig{\ref{figure:AC_map}}(a), the power spectrum ($|\rm{FFT}(\Delta A)|^2$) with only the phonon-induced oscillations shows a maximum intensity near the Fermi level of 40.4\,eV, while above 42\,eV, weaker features are associated with the appearance of phonon-induced features from the large lattice distortion found in \Fig{\ref{figure:spectrum1}}(b).

\begin{figure*}[t!]
 \centering
\includegraphics[width=\textwidth]{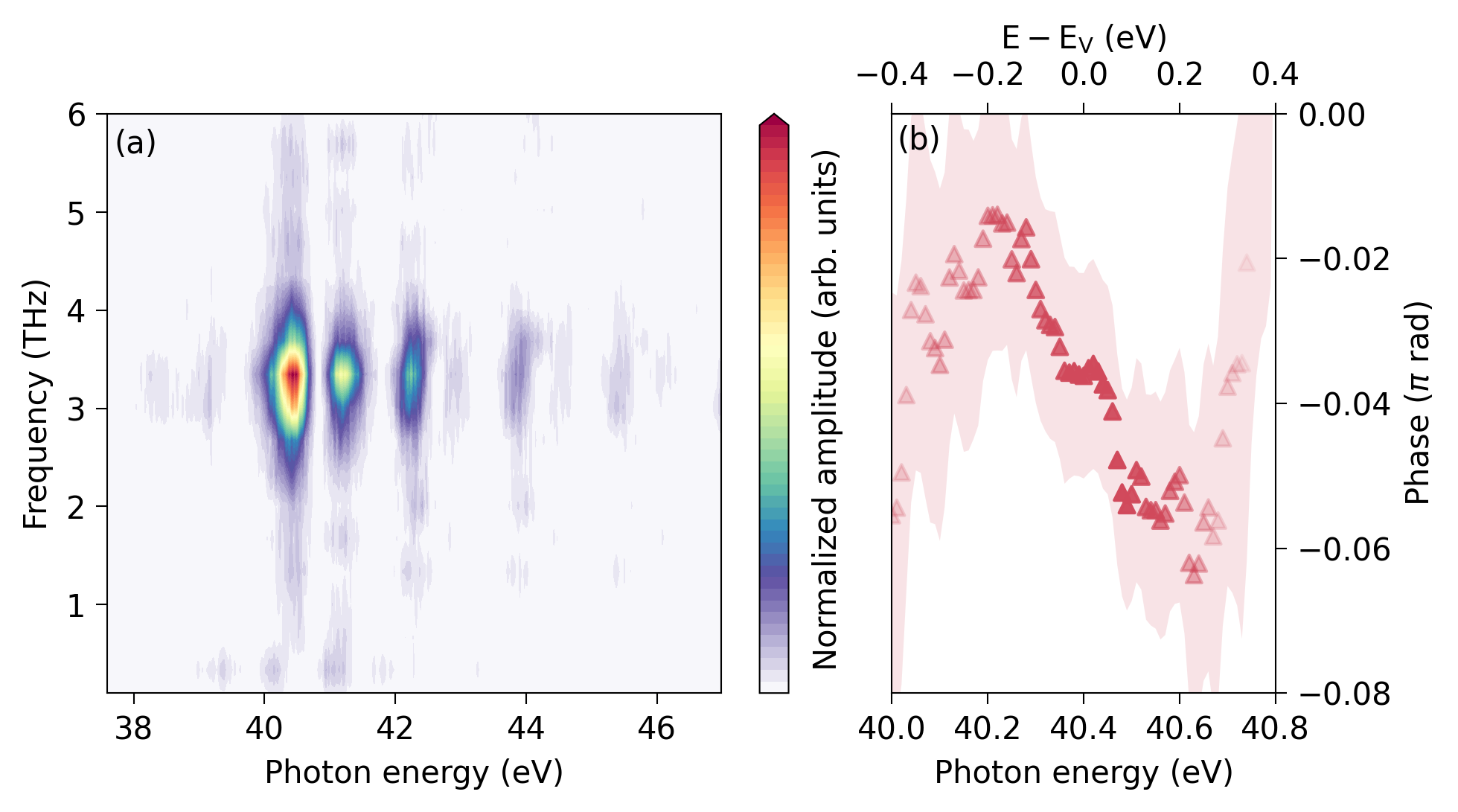}
\caption{Fourier analysis of the coherent phonon motion is performed by subtracting the non-oscillatory time dependence from the SVD. (a) The contour of the power spectrum. The energy resolved phase of the phonon motion near the Fermi level is shown in (b) at 3.17 THz retrieved via a non-linear least squares fit using the parameters from the multi-temperature model. This corresponds to the softened \Ag{} phonon mode frequency.}
\label{figure:AC_map} 
\end{figure*}

In \Fig{\ref{figure:AC_map}}(b), the spectral phase is shown for the \Ag{} mode in vicinity of the Fermi level. In agreement with the phase of $-0.04(1)\pi$\,rad retrieved from the multi-temperature model, the average phase of the \Ag{} phonon from Fourier analysis in \Fig{\ref{figure:AC_map}}(b) at the Fermi level is $-0.05(2)\pi$\,rad. The phase observed here is in good agreement with the previously reported value of $-0.04(5)\pi$\,rad \cite{ZeigerPRB1992}. It is however worth noting that there is a non-flat variation of the spectral phase across the XUV absorption energy. In the case of a purely displacive excitation, the phase of the \Ag{} phonon should be zero with a small phase imparted by the dampening \cite{ZeigerPRB1992}. The dampening induced phase shift calculated from the phonon dampening and carrier relaxation is found to be less than $0.01\pi$\,rad. O\'Mahony \cite{MahoneyThesis} demonstrated in the A7 semimetals that there is a dependence of the phonon phase on lifetime of the electronic excitation, which results in a mixing of a displacive excitation force (flat phase) and an impulsive force that depends on carrier relaxation. The XUV energy dependence of the \Ag{} phonon phase near the Fermi level in \Fig{\ref{figure:AC_map}}(b) thus could be due to such an energy dependent electron-phonon coupling \cite{MahoneyThesis}. In the case of a pure displacive excitation, where there is no electronic excitation lifetime dependence, the force exerted on the \Ag{} phonon mode will be constant over the pump pulse and there should be no energy-dependence of the phase. In Te, the measured phase exhibits a small but observable, $0.05\pi$\,rad variation. This implies a small, but non-zero effect of carrier relaxation on the \Ag{} phonon driving force. Despite the hot carrier energy thermalization through electron-phonon scattering, which occurs over $T_{ep} = $ 196(8)\,fs, this appears to have little impact on the phase. When analyzing the character of the bands excited by the 1.3-2.5\,eV pump pulse, this is not surprising. The bandwidth covers the non-bonding $5p$ lone pair (-3\,eV to VBM) and the anti-bonding $5p$ orbitals (CBM to 3\,eV above) crucial to the helical chain structure \cite{Kirchoff1994}. All the excited electrons are thus likely to cause a weakening of the intra-chain bonds, as opposed to a strong energy-dependent contribution, which would therefore produce a near pure displacive excitation. This could be tested in detail with theoretical calculations of the energy dependent carrier relaxation as has been done for 2D tellurium \cite{Xia2024}, or by calculating the band dependence of the force along the \Ag{} phonon coordinate \cite{MahoneyThesis}.

While we found that the \Ag{} coherent phonon and hot carriers are not responsive to changes in laser polarization in the absorption detection geometry with these polycrystalline samples, future studies on samples oriented perpendicular to the c-axis could enable measurement of the non-totally symmetric modes through the photo-Dember effect \cite{DekorsyPRB1996}. It may be interesting to explore whether the \Eg{} mode or other phonon modes may couple to the electronic temperature or influence motion along the \Ag{} coordinate.

\section{Summary}

Using few-femtosecond core-level transient absorption at the \TeN{} edge we have observed a coherent coupling of the photo-excited holes and electrons to the \Ag{} phonon mode in elemental tellurium. Following excitation of the $5p$ lone-pair orbitals to the anti-bonding conduction band, the lattice is excited along the \Ag{} phonon coordinate to a new excited state that lives on the timescale of carrier recombination. This excited lattice configuration is found to follow the 33(5)\,ps carrier population decay, while hot carrier relaxation ($\Delta T_e$) driven by electron-phonon coupling is much faster and occurs over 1.59(3)\,ps. Coherent excitation of the \Ag{} phonon mode modulates the hot carrier temperature ($\Delta T_e$), in which when the phonon displacement is maximized (Te more symmetric), the carrier temperature is minimized due to an increase in electronic heat capacity as Te becomes more metallic and the DOS is increased at the Fermi level. This back-coupling of the lattice energy to the carriers is observed for the first time in a semiconductor and is driven by a coupling between the electronic and lattice degrees of freedom along the \Ag{} coordinate. These results that demonstrate a dependence of the electronic DOS on $\Delta Q_0$ are consistent with a coherent phonon driven tunability of the electronic structure \cite{Ning2022,Roeser2002,Kim2003}. While depopulating the lone-pairs and populating the anti-bonding conduction band leads to the excited state Te structure, Fourier analysis demonstrates the launching of the coherent phonon is not through a pure displacive excitation, but has a weak dependence upon carrier relaxation possibly due to energy-dependent electron-phonon coupling. The measurement of coherent coupling between the hot carriers and \Ag{} phonon mode in elemental Te provides the basis for further investigation of the light-induced increase in metallicity and how it may couple to other degrees of freedom, such as spin or the non-totally symmetric phonon modes.

\section{Acknowledgments}

This work was supported by the U.S. Department of Energy, Office of Science, Basic Energy Science (BES), Materials Sciences and Engineering Division under contract DE-AC02-05CH11231 within the Fundamentals of Semiconductor Nanowires Program (KCPY23) for performance of the extreme ultraviolet laser experiments, preparation of samples, and analysis and interpretation of the results. Support for laser instrumentation and vacuum hardware is from AFOSR grant numbers FA9550-19-1-0314, FA9550-24-1-0184, and FA9550-22-1-0451. This material is based upon work supported by the National Science Foundation Graduate Research Fellowship Program (NSF GRFP) under Grant No. DGE 2146752 and DGE 1752814. Any opinions, findings, and conclusions or recommendations expressed in this material are those of the author(s) and do not necessarily reflect the views of the National Science Foundation. J.R.A and H.K.D.L acknowledge support from the NSF GRFP. H.L. acknowledges support from the Swedish Research Council (2023-06502) and the Sweden-America Foundation, L.D. acknowledges the European Union’s Horizon research and innovation programme under the Marie Sklodowska-Curie grant agreement No 101066334—SR-XTRS-2DLayMat.

\appendix

\section{Experimental Apparatus and Data Processing}

The XUV transient absorption experiments are conducted on a table-top apparatus that starts with a Ti:sapphire laser (Coherent Astrella). The output is 7\,mJ at 1\,kHz with a pulse duration of 35\,fs at a center wavelength of 795\,nm. From the laser output, 2.3\,mJ is reflected from a beam splitter and focused into a 2.2 meter long 400 micron inner diameter hollow core fiber filled with neon at a pressure of 30\,Psi for supercontinuum generation. The supercontinuum output of the hollow core fiber spans an octave from 500 to 1000\,nm with a pulse energy of 1.5\,mJ and is mechanically chopped to 100\,Hz prior to collimation. The beam is subsequently collimated with a magnification of 1.25 to the fiber input and compressed with seven pairs of PC70 double angle chirp mirrors (Ultrafast Innovations) that support the entire beam bandwidth and a Z-cut 2\,mm thick ammonium dihydrogen phosphate crystal \cite{Timmers2017}. The resulting beam is sent into the Mach-Zender interferometer via a 1\,mm thick beam splitter (90T:10R), where the transmitted beam is used for HHG, and the reflected portion is used for the pump. Both pump and HHG driving pulse are optimized with fused silica wedges for fine tuning of the dispersion and are approximately 4\,fs in duration. The HHG driving arm is focused (50 cm focal length) into an krypton filled gas cell for HHG where a broadband XUV (35-73\,eV) pulse is generated as the probe pulse. The residual NIR-VIS driving field is filtered out with a 100\,nm thick Al foil that acts as a high-pass filter. The XUV pulse is focused onto the thin film sample by a gold-coated toroidal mirror implemented in a 2f-2f imaging configuration and subsequently dispersed onto a CCD camera (PIXIS 400B) in a flat-field grating spectrograph where the transmitted spectrum is measured. The pump in the Mach-Zender is time delayed to the probe by an optical delay line with an optical encoder (Attocube), and focused (1\,m focal length) onto the sample after recombining in a collinear geometry with the XUV by an annular mirror. The pump beam is removed by a second 100\,nm thick Al foil prior to entering the spectrograph chamber. The pump beam is both transmitted and blocked at each time delay to collect XUV spectrum with and without the pump. 

The TA spectra are calculated at each time delay as $\Delta A = \log_{10}(I_{\rm{on}}/I_{\rm{off}})$, where $I_{\rm{on}}$ and $I_{\rm{off}}$ are the transmitted XUV with the NIR pump on and off, respectively. The XUV TA spectra are subject to de-noising via an edge-referencing scheme where no TA signal is observed to improve the noise floor to about 1\,mOD  \cite{Geneaux2021OE}. This is done for a total of 93 cycles, over 465 time delays. Included in \Fig{\ref{figure:raw_data}}(a) is the de-noised TA without subtraction of the pre-zero delay heat background. To subtract the time-independent heat background, the first five negative time delays are averaged in time and subtracted from the TA spectra. The raw unprocessed data is shown in \Fig{\ref{figure:raw_data}}(b), where there is prominent high frequency noise.

\begin{figure}[!htb]
\begin{center}
\includegraphics[width=\columnwidth]{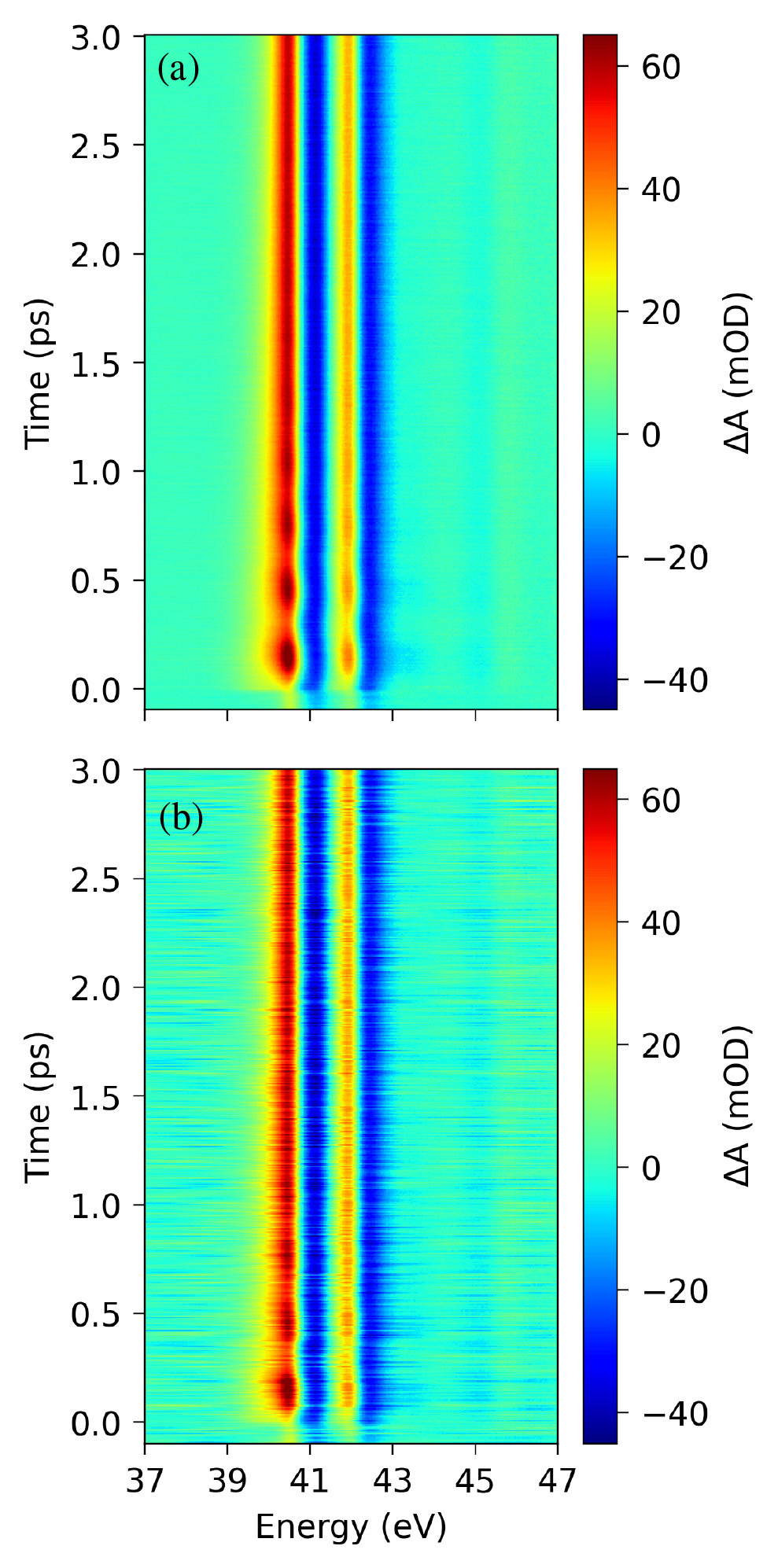}
\caption{ (a) Processed TA data with a visible pre-time-zero delay signal from static heat accumulation. (b) Raw unprocessed TA data where the oscillation pattern is the prominent spectral feature.}
\label{figure:raw_data} 
\end{center}
\end{figure}

\label{instrument}

\section{Film characterization}

In \Fig{\ref{figure:AFM}} we present an atomic force microscopy image of the Te thin film used in the experiment. Over a large area the film is found to exhibit uniform coverage, with a surface roughness on the order of a few nanometers. Additionally, part of the substrate was masked during the deposition enabling no film to be deposited on part of the substrate. From this, the film is estimated to be on the order of 100\,nm. Further, the crystallinity is identified by powder x-ray diffraction of a film on the SiN substrate used in the pump-probe where the peaks are labeled by material via their miller indices in \Fig{\ref{figure:XRD}}. The identifiable miller indices of trigional Te (RRUFF ID: R070376.1) \cite{Lafuente2016} confirm that the Te film is polycrystalline as expected.

\begin{figure}[!htb]
\begin{center}
\includegraphics[width=\columnwidth]{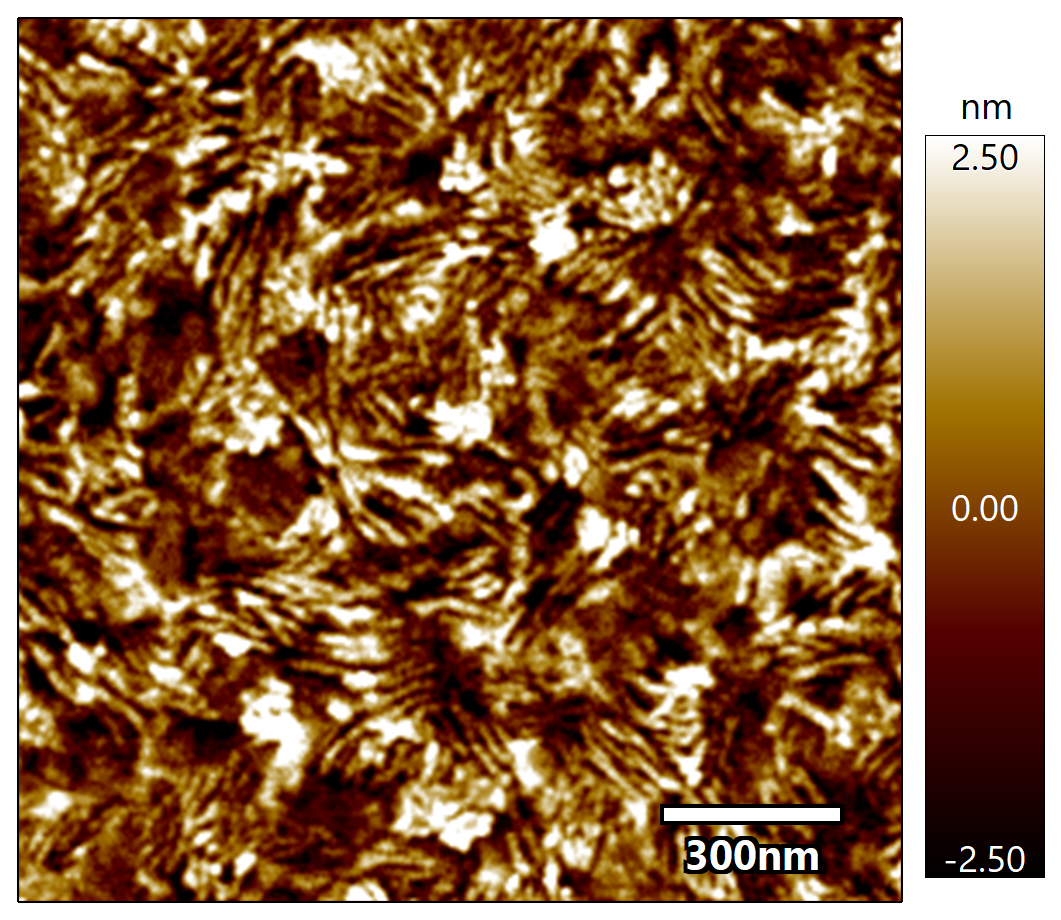}
\caption{AFM image of the Te films}
\label{figure:AFM} 
\end{center}
\end{figure}

\begin{figure}[!htb]
\begin{center}
\includegraphics[width=\columnwidth]{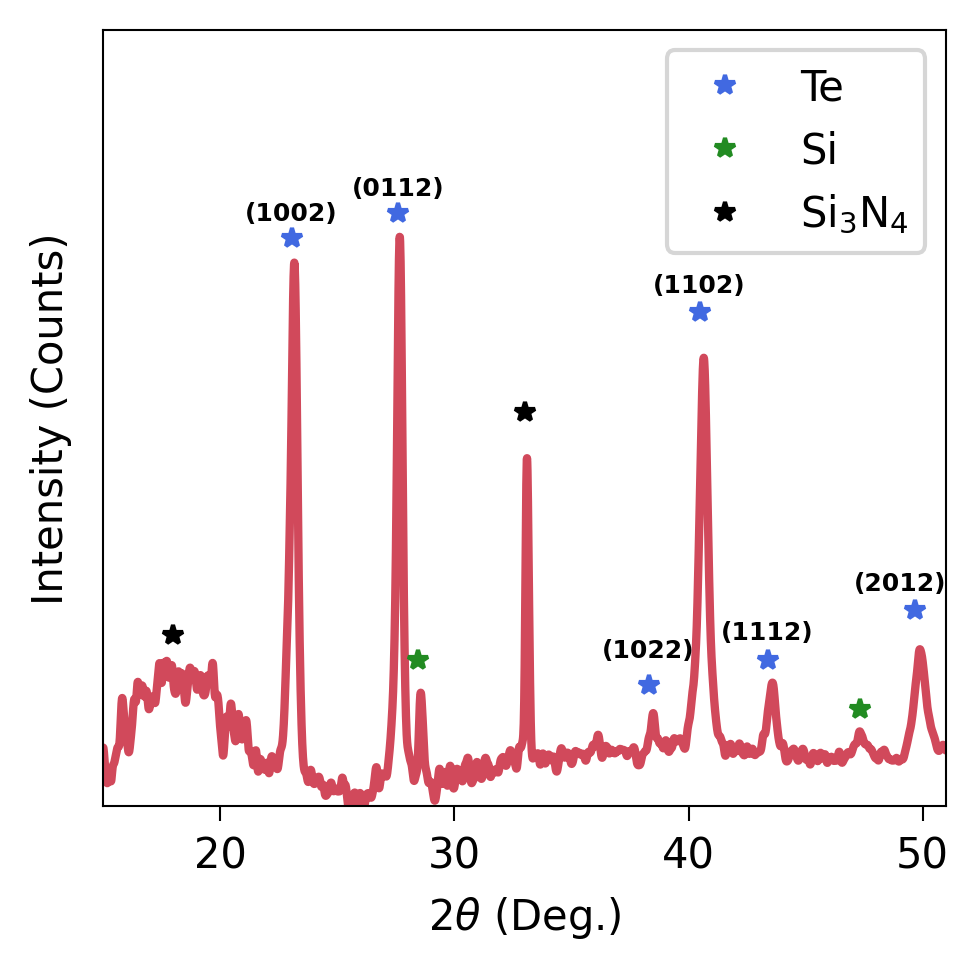}
\caption{Powder diffraction pattern of the Te thin film on a SiN/Si substrate. The labels of each material correspond to identified miller indices, while the specific indices are included for tellurium.}
\label{figure:XRD} 
\end{center}
\end{figure}

\section{Instrument response function}
\label{IRF}

In order to characterize the instrument response time, we measure the singly excited $2ssp^6(^2S)3p^1P^0$ Rydberg resonance in neon located at 45.55\,eV \cite{Schulz1996}. Unlike the above measurements where the XUV spectrum is measured after interaction with the NIR and there is only a transient signal after time zero, in neon the XUV comes before the NIR. This enables the NIR to perturb the oscillating XUV dipole near the resonance by coupling to nearby states and the continuum producing a line-shape change and suppression useful to characterize few-cycle pulses. \Fig{\ref{figure:cross}}(a) demonstrates the change in XUV intensity at the n=3 resonance due to the NIR. Using SVD analysis and analyzing the temporal vector of the first component, we retrieve a 6.4(5)\,fs instrument response time by a least squares fit with an exponentially modified Gaussian shown in \Fig{\ref{figure:cross}}(b). At the start of each measurement this methodology is used to optimize the pump pulse contrast and duration by fine tuning of the dispersion. The neon $2ssp^6(^2S)np^1P^0$ Rydberg resonances are additionally used to calibrate the spectrometer. 

\begin{figure*}[t!]
 \centering
\includegraphics[width=\textwidth]{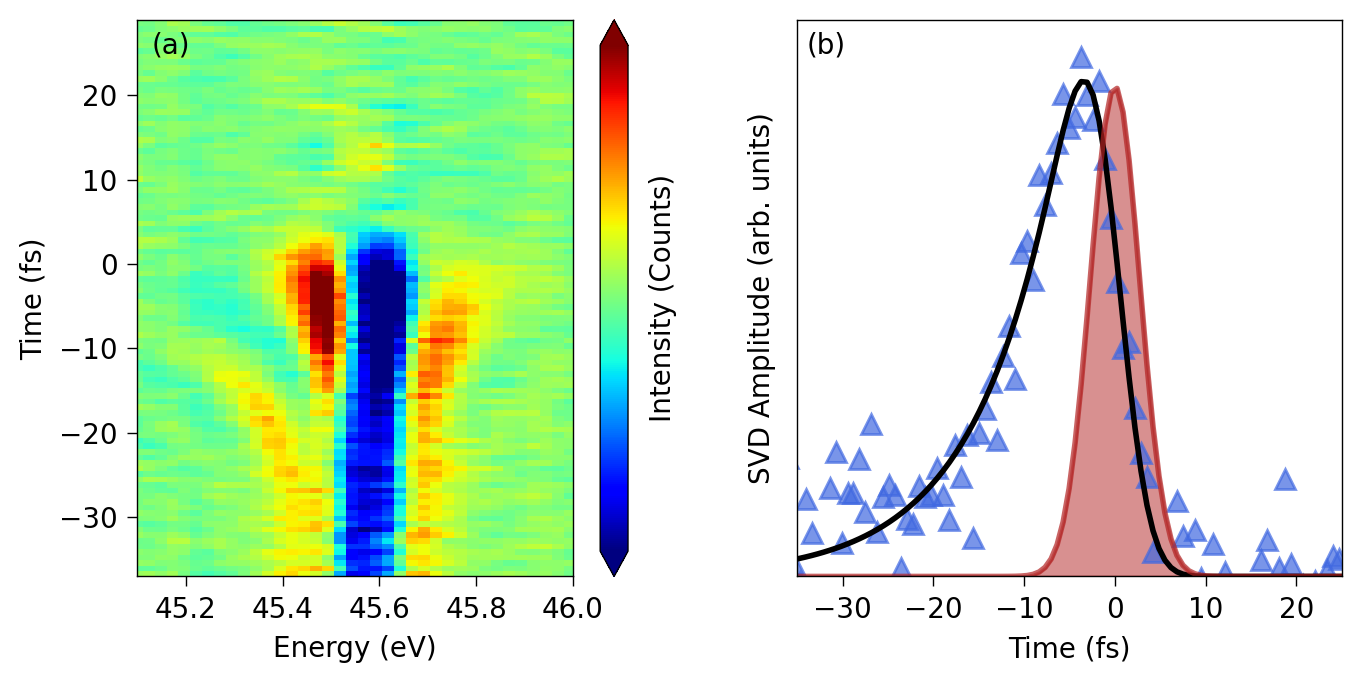}
\caption{(a) The change in counts at the $2ssp^6(^2S)3p^1P^0$ Rydberg resonance in neon. (b) The time dependence of the 1st singular value component from an all pump-on measurement of the neon Rydberg resonance at 45.55\,eV. The black curve is an exponentially modified Gaussian fit with the cross-correlation as a free parameter.}
\label{figure:cross} 
\end{figure*}

\section{Analytical Decomposition}
\label{Appendix:decomp}

We analyze here using an analytical model predominant spectral features in the transient absorption spectra at 150\,fs, corresponding to the first oscillation maximum of the phonon. Well demonstrated in previous core-level TA measurements of elemental semiconductors \cite{Zurch2017} and transition metal dichacolgenides \cite{Attar2020,Oh2023}, the excited state can be decomposed into individual contributions:
\begin{equation}
\label{eq:simple}
\Delta A = \Delta A_{\rm{Shift}}+\Delta A_{\rm{Carriers}}+\Delta A_{\rm{Broadening}}
\end{equation} 
incorporating the phonon and band-gap renormalization induced absorption edge shift ($\Delta A_{\rm{Shift}}$), excited state inhomogeneous broadening of the core-level transition ($\Delta A_{\rm{Broadening}}$), and carrier induced state blocking (filling) in the conduction (valence) band ($\Delta A_{\rm{Carriers}}$), also called state removal/state opening. In the region of the \TeN{} edge we show the decomposition into these components in \Fig{\ref{figure:model}}. To retrieve the individual contributions, an optimization is performed at 1\,eV around the Fermi level and the fitting errors retrieved via the Hessian matrix.

In this analytical model the absorption edge shift is optimized to a 580(30)\,meV uniform energy shift (rigid shift approximation) with respect to the ground state XUV absorption. An optimized 8.3(4)\,\% scaling factor is applied to the ground state XUV absorption to obtain agreement with the $\Delta$A. We believe the scaling factor is due to non-uniform shifting and restructuring of the conduction band DOS across the BZ following photoexcitation \cite{Zurch2017, Geneaux2021,Tangney2002}. The shift includes both changes to the conduction band energy and 4d level binding energies that we cannot independently discern. The inhomogeneous broadening of the excited state is achieved by convolving the static absorption with a Gaussian and yields a broadening $\sigma=40(10)$\,meV. As hot carrier thermalization is expected to occur mostly within the first 100\,fs, we approximate the carrier induced state filling and blocking as a Fermi-Dirac distribution over the DOS \cite{Jain2013} with 0.5 eV lifetime broadening. This is replicated for the spin-orbit split core-levels where there is strong spectral overlap between the holes of the $4\rm{d}_{3/2}$ and electrons of the $4\rm{d}_{5/2}$. From this, we estimate a shared electronic temperature of 3300(300)\,K for both holes and electrons at the maximum of the phonon displacement. Due to spectral congestion at the electron state blocking signal, the assumption is made of a shared electronic temperature. Additionally, the quasi-Fermi levels of the carriers are offset by half the energy of a pump photon (0.8 eV) from the ground state Fermi level \cite{Zurch2017}. These approximations are made to reduce the over parametrization of the optimization. However, this analytical model slightly over parametrizes the data, with a $\chi_\nu^2=0.94$, but has a benefit of isolating the main spectral features.

While reproducing the change in absorption near the Fermi level, the rigid shift approximation made in the model causes deviations from experiment due to an overestimate of the edge shift at high energies (above 42\,eV). This we attribute to a non-uniform electron-phonon coupling over a large photon energy range, as not all bands renormalize equally \cite{Zurch2017, Geneaux2021}. In addition to the rigid shift approximation, the use of a scaling factor in the model warrants further consideration. When considering how the conduction band DOS responds to the \Ag{} phonon induced lattice distortion, it is not a rigid red shift of all conduction band DOS, but the appearance of a small component towards the Fermi level while depleting the near CBM DOS. There is also considerable restructuring over the entire DOS \cite{Tangney2002}. This is in contrast to changes to the DOS in Bi upon a 1\% increase in inter-atomic distances \cite{Geneaux2021}, where there is limited restructuring and it resembles a linear red shift of the conduction band in the vicinity of the Fermi level. In the case of Te, particularly under large lattice distortions, it is thus no surprise that the rigid shift approximation in the absence of a scaling factor breaks down. The scaling factor would suggest only a fraction of the DOS undergoes shifting in response to photoexcitation, consistent with how the DOS evolves with large lattice distortions \cite{Tangney2002}. The 4d core-level potential should, however, follow the traditional methodology of a rigid shift. Future investigations might consider a more advanced methodology, dependent upon detailed TDDFT calculations of how the DOS evolve with increasing lattice distortion, local field effects \cite{Schumacher2023}, and the core-level potential slope, and their changes can be approached in an independent manner.

Although this approach provides precise quantities within the approximations, lack of detailed knowledge of complex band dynamics and strong spectral overlap of lattice and carrier features gives rise to significant uncertainty. Further the necessity to utilize a scaling factor in the rigid shift approximation due to an unknown dependence of the shift on band dispersion, energy, and DOS restructuring impedes successful interpretations with this decomposition method. 

\begin{figure}[!htb]
\begin{center}
\includegraphics[width=\columnwidth]{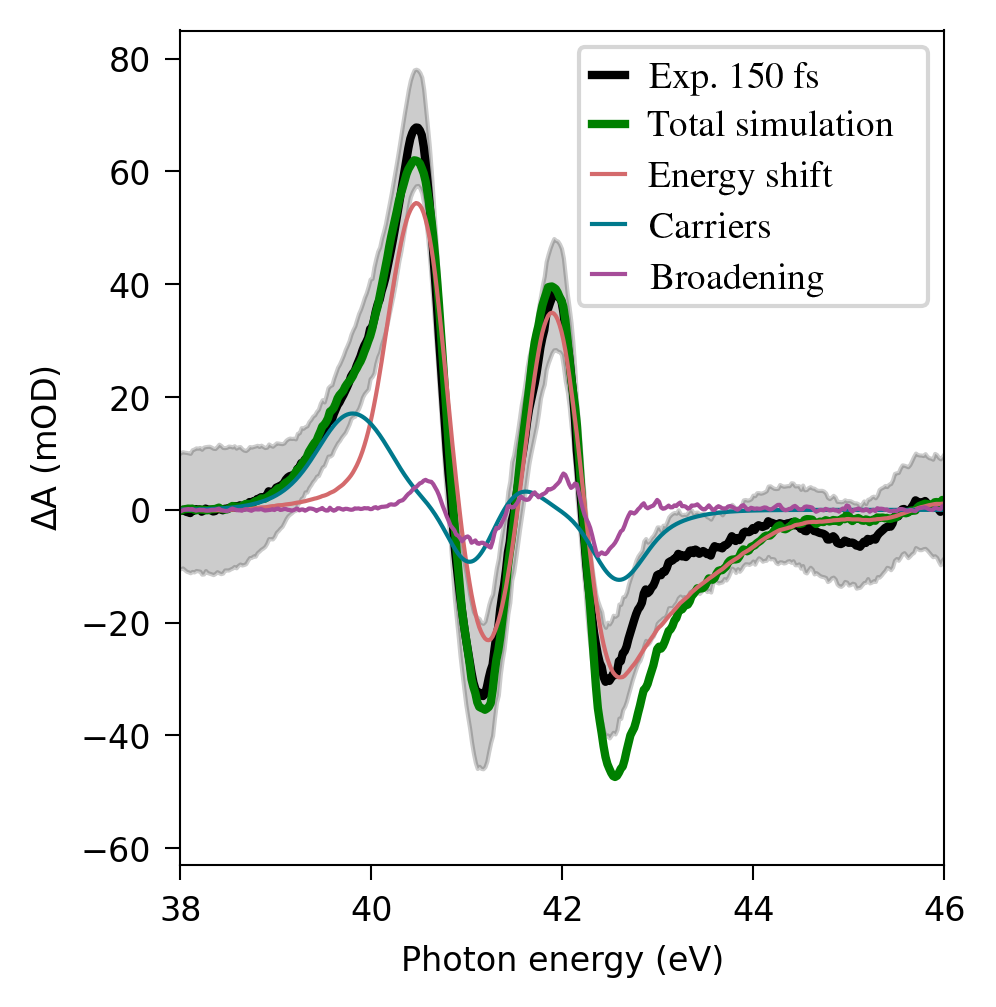}
\caption{The spectral decomposition of the TA spectrum at a pump-probe delay around 150\,fs by the deconvolved components of shifting, broadening, and state filling/blocking. The black curve corresponds to the measured $\Delta A$, while the green curve is the fit with the analytical model of shifting, broadening and carrier state filling/blocking. The energy shift (red), broadening (purple), and carriers (blue) are shown as distinct contributions to the TA. The shaded grey region represents an upper bound of the statistical error retrieved from the spread of the unprocessed data.}   
\label{figure:model} 
\end{center}
\end{figure}

\section{Residual Transient Absorption}

The residual TA signal is presented in \Fig{\ref{figure:resdiual}} where there is a few mOD signal near the 4d$_{5/2}$ to Fermi level (40.4\,eV) transition that rapidly decays on the timescale of a couple hundred femtoseconds. This residual increase in absorption is well below the Fermi level, and a decrease near the Fermi level is consistent with non-thermalized holes and electrons not captured by the SVD that we assign to the hot carriers. While two-photon absorption could occur producing non-thermal holes, no increase in absorption discernible from the noise level is observed below below 38.5\,eV, which exceeds the single photon energy of the NIR pump minus the bandgap. Additionally, the hot carriers dissipate their energy into low energy carriers near the Fermi level that persist to 33\,ps, which results in the appearance of state blocking/filling contributions emerging in the residual TA after the 1.6\,ps decay. In addition to evolution of the state blocking via the hot carriers, the phase variation in energy of the phonon motion is not possible to capture with a single SVD component and is thus found in the residual TA. This phase variation in energy of the \Ag{} phonon motion results in oscillatory features in the residual TA. 

\label{Appendix:Residual}

\begin{figure}[!htb]
\begin{center}
\includegraphics[width=\columnwidth]{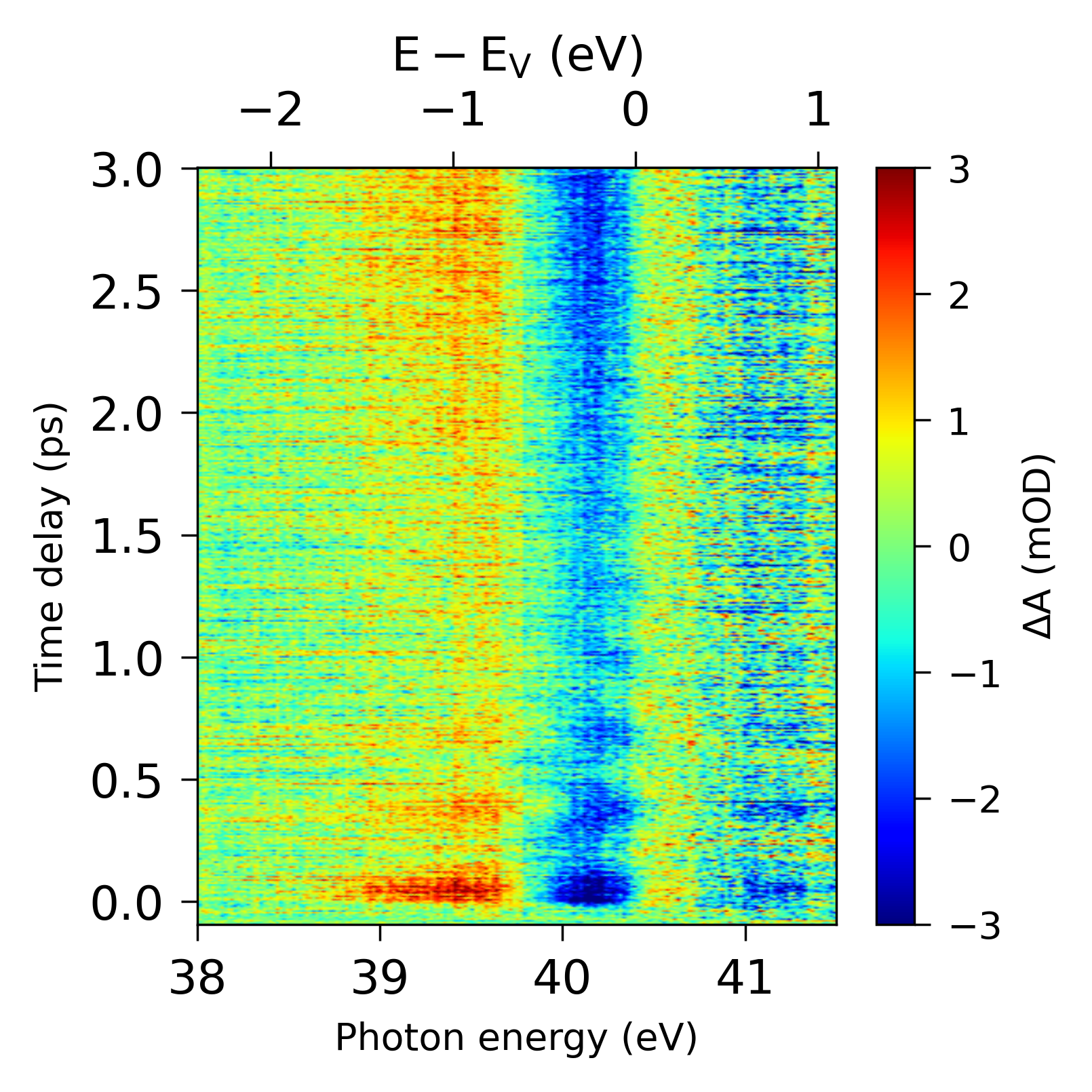}
\caption{The residual transient absorption signal after removing the three largest singular value components from the overall signal. At timescales less than 1.5\,ps, oscillations are visible due to the energy dependence of the coherent phonon phase not captured by the singular value vectors analyzed. Additional contributions of non-thermalized carrier distributions are expected within the first few hundred fs, while beyond 2\,ps the temporal dependence of the carrier temperature is visible. }
\label{figure:resdiual} 
\end{center}
\end{figure}

\section{Low fluence measurement}

In \Fig{\ref{figure:low_power}} we present the transient absorption spectra at a carrier density of $2\cdot$\cd{d20}. At this lower carrier density, below the expected threshold for the photo-induced semiconductor-to-metal phase transition, the phonon induced edge shift is clearly observed, however, the substantially smaller state blocking/filling contributions are not resolved. The SVD applied to these measurements reveals a single SVD vector resembling the phonon and electronic screening induced absorption edge shift. All other SVD components are indistinguishable from noise. From this, we find no measurable decay ($\kappa$) of the excited state equilibrium position and coherent phonon ($\gamma$) on the timescale of the measurement. The retrieved frequency is 3.6(3)\,THz, closer to the equilibrium frequency due to a reduced electronic softening. 

\label{appendix:low_fluence}
\begin{figure}[!htb]
\begin{center}
\includegraphics[width=\columnwidth]{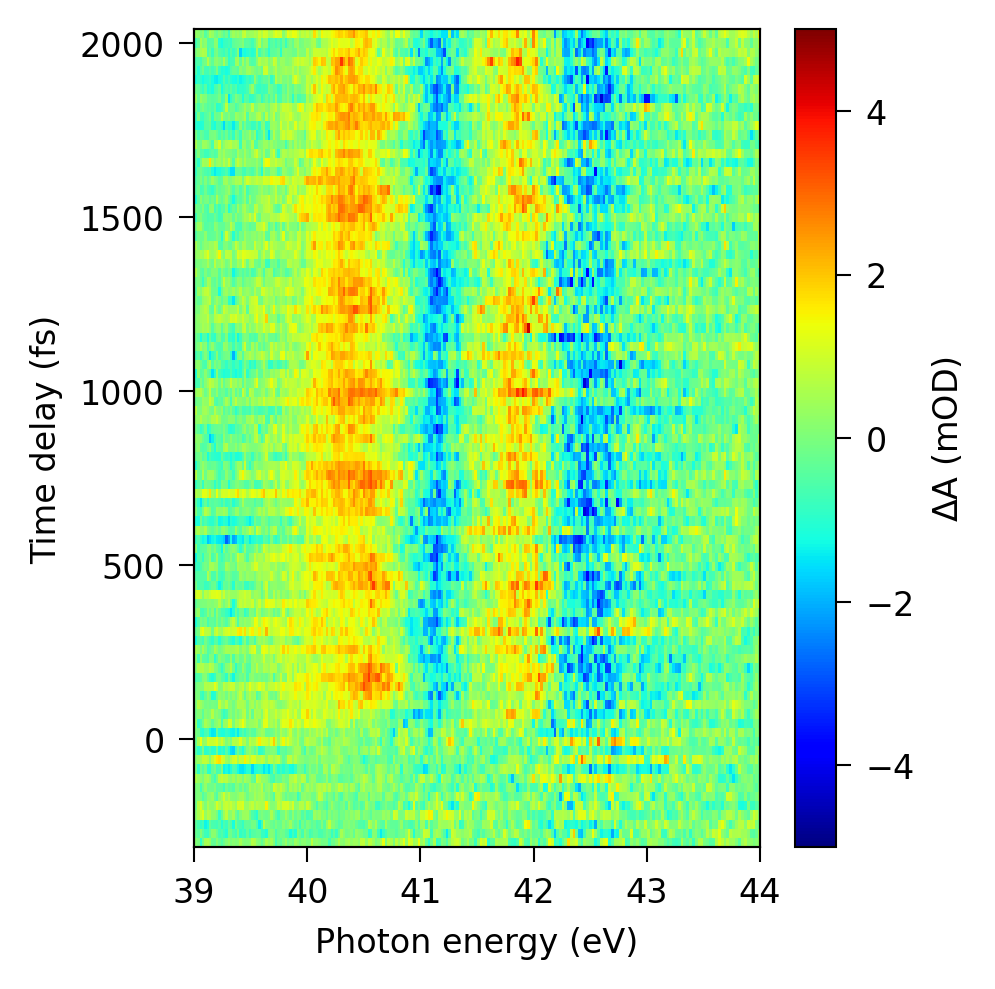}
\caption{Measured XUV transient absorption spectra at the \TeN{} edge between -300 and 2000\,fs at a carrier density of $2\cdot$ \cd{d20}. The time axis corresponds to the arrival of the XUV probe relative to the optical pump arriving at time zero. Photoexcitation of Te with the pump launches a coherent phonon motion that appears as observed oscillation pattern in the differential absorption.}
\label{figure:low_power} 
\end{center}
\end{figure}

\section{Short timescales}

As discussed above in section \ref{sec:discussion}, there is an early timescale rise of the energy shift trace that is too fast to be explained by coherent phonon motion. In \Fig{\ref{figure:early_scan}} we show the energy shift trace component of the SVD analysis to a measurement utilizing 660\,as time-steps, and the TA signal integrated over the energy ranges 40.35-40.5\,eV and 38.5-40\,eV. In conjunction with the Te measurement, TA of the neon Rydberg resonances are done after each scan cycle to correct for zero-delay drifts, where the standard deviation of zero-delay drifts over 113 scan cycles is found to be 508\,as, indicating minimal long-term drifts. The energy range of 40.35-40.5\,eV can be mostly attributed to the absorption edge energy shift contribution based on the analytical model (see Appendix \ref{Appendix:decomp}), while the 38.5-40\,eV range is the holes. We find the rapid rise of the energy shift vector is near instantaneous with the cross-correlation, suggesting a rapid electronic response that contributes to a red shift of the \TeN{} absorption edge, followed by the start of a slow-rise due to the onset of phonon motion. This is consistent with the energy integrated TA signal over the 40.35-40.5\,eV region. The energy integrated TA signal over the hole region (38.5-40\,eV) exhibits a rise that follows the cross-correlation, but maintains a near constant integrated signal outside of the cross-correlation on these short timescales. Despite the pump field intensity being on the order of the approach to non-linear effects - in semiconductors such as Ge \cite{Inzani2023} and GaAs, \cite{Schlaepfer2018} non-linear excitation has been observed via XUV - we do not find in the integrated TA signal over the hole region any dynamics associated with non-linear excitation effects in the measurements. This outcome could be due to attosecond timing instability during an individual scan cycle and there may be a need to sample at a higher frequency. Additionally, the weak pre-time-zero delay signal outside the cross-correlation is attributable to a weak pre-pulse in the pump beam, visible at around 10\,fs in \Fig{\ref{figure:cross}}(a). 

\label{section:early}

\begin{figure}[!htb]
\begin{center}
\includegraphics[width=\columnwidth]{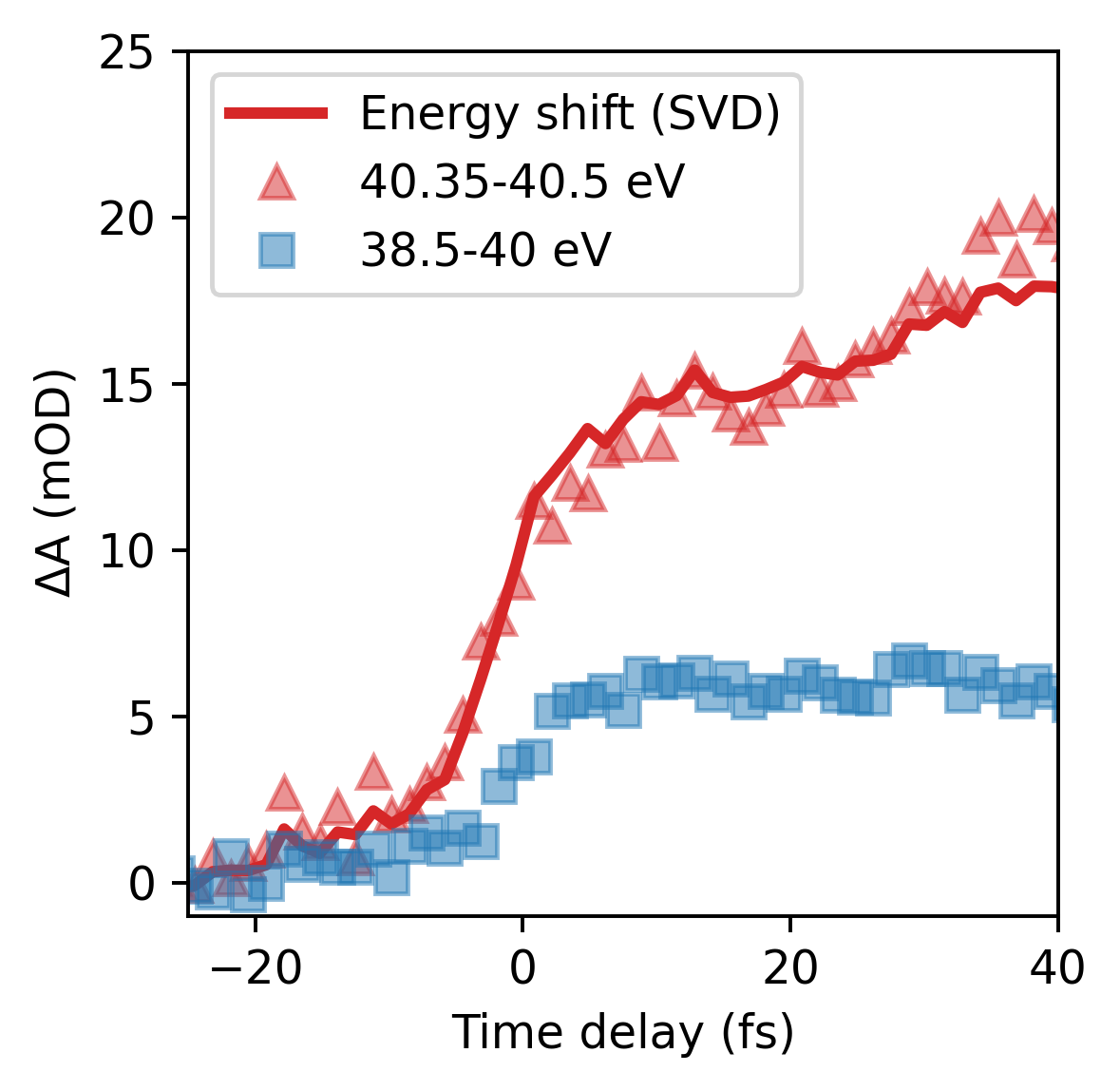}
\caption{The energy shift vector retrieved from the 1st SVD component at a carrier density of 1.5 $\cdot$ \cd{d21} measured on a shorter timescale than presented in the main text. Also included are traces of the TA signal retrieved by signal integration over the energy ranges 40.35-40.5\,eV and 38.5-40\,eV.} 
\label{figure:early_scan} 
\end{center}
\end{figure}

\bibliography{references.bib}

\end{document}